\documentclass[11pt]{JHEP}
\usepackage{amsmath}
\usepackage{amssymb}
\usepackage{amsmath,amsthm,epsfig,euscript,array,cancel}
\font\mybb=msbm10 at 10pt
\def\bb#1{\hbox{\mybb#1}}

\newcommand{\be}{\begin{equation}}
\newcommand{\ee}{\end{equation}}

\def\bea{\begin{eqnarray}}
\def\eea{\end{eqnarray}}
\preprint{ 
October 2,  2018. V3: Misprints corrected 19/03/22 }
\renewcommand{\theequation}{\arabic{section}.\arabic{equation}}
\title{Supersymmetric action for multiple D$0$-brane system  }

\author{Igor Bandos
 ~ \\ {\small\it UPV/EHU, P.O. Box 644, 48080 Bilbao, Spain}  ~\\ and ~ \\ {\small\it
IKERBASQUE, Basque Foundation for Science, 48011, Bilbao, Spain}
}

\date{21/08/2018,  Printed \today }

\abstract{We have constructed a complete action for the system of $N$ D$0$-branes in flat 10D type IIA   superspace. It is invariant under the rigid spacetime supersymmetry and local worldline supersymmetry ($\kappa$--symmetry). This latter can be considered as supersymmetry of maximal 1d $SU(N)$ SYM model which is made local by coupling to supergravity induced by embedding of the center of energy worldline into the  target superspace. The spinor moving frame technique is essentially used to achieve such a coupling.
We discuss the differences with Panda-Sorokin multiple $0$-brane action and with the dimensionally  reduced 11D multiple M-wave action.
}

\keywords{}

\begin{document}

\section{Introduction}

In 1995 E. Witten agrued
\cite{Witten:1995im} that the system of N nearly coincident D$p$-branes carries non-Abelian gauge fields on a center of mass worldvolume  and that at very low energy it is described by the action of $U(N)$ maximally supersymmetric Yang--Mills (SYM) theory. In it the $U(1)$ sector describes the center of mass motion of the multiple D$p$-brane (mD$p$) system while the $SU(N)$ sector describes the relative motion of the mD$p$  constituents. Actually, $U(1)$ SYM action decouples and can be identified as a low energy limit of gauge fixed version of the complete nonlinear action for single D$p$-brane
\cite{Cederwall:1996pv,Cederwall:1996ri,Aganagic:1996pe,Bergshoeff:1996tu,Aganagic:1996nn,Bandos:1997rq}.

Then the natural problem was to  find a complete action for multiple D$p$-brane system. It was approached in a number of papers and certain progress was reached during these years  \cite{Tseytlin:1997csa,Myers:1999ps,YLozano+=0207,Lozano:2005kf,Howe+Lindstrom+Linus}. {In particular, the bosonic limit is widely believed to be given by the  Myers's `dielectric brane' action}  \cite{Myers:1999ps} which was obtained from the requirement of consistency with T-duality transformations of D-branes and background fields.
A very interesting construction on '-1 quantization level' was proposed in  \cite{Howe+Lindstrom+Linus}. There such a dynamical system was constructed, that its  quantization should reproduce the desired multiple D$p$-brane ({\it mD$p$}) action. However, the complete realization of this step in a fool glory seems to imply the quantization of the complete interacting system of supergravity and super-Dp-brane.

A complete action including fermions and invariant under spacetime supersymmetry and local fermionic $\kappa$-symmetry  is known for the system of ten-dimensional ({\it 10D})  multiple $0$-branes
\cite{Sorokin:2001av,Panda:2003dj} as well as for 11D  multiple M0 ({\it mM$0$} or multiple M-waves) system
\cite{Bandos:2012jz,Bandos:2013uoa}. Besides these, in D=3 some complete ${\cal N}=1$ supersymmetric multibrane actions are known \cite{Drummond:2002kg,Bandos:2013swa}.
 Furthermore,  the infrared fixed points of the system of N M2-branes is believed to be described by   Bagger--Lambert--Gustavsson (BLG) model
\cite{Bagger:2006sk,Gustavsson:2007vu,Bagger:2012jb} for N=2 and by  Aharony--Bergman--Jafferis--Maldacena (ABJM) model \cite{Aharony:2008ug,Bagger:2012jb} for $N\geq 2$. The infrared fixed point of  multiple M$5$-brane system should reproduce  an enigmatic $D=6$ $(2,0)$ superconformal theory; recently it was conjectured \cite{Douglas:2010iu,Lambert:2010iw} that this can be described by $D=5$ SYM model.

The mM$0$ dynamical system of \cite{Bandos:2012jz,Bandos:2013uoa} can be considered as 11D massless superparticle carrying on its worldline 1d ${\cal N}=16$ $SU(N)$ SYM multiplet.
It was natural to expect  that the dimensional reduction of mM$0$ action should reproduce a multiple D$0$-brane (mD$0$) action.
Surprisingly the result of such a dimensional reduction looks quite complicated  and does not resemble what we
expected for the mD$0$-brane action (we discuss this problem in the Appendix C).
This is why in this paper we construct a supersymmetric 10D multiple D$0$-brane action with local fermionic kappa-symmetry directly, putting maximally supersymmetric 1d SYM multiplet on the effective worldline of a center of mass of the mD$0$-system and coupling it to the induced worldline supergravity. We also discuss the differences of our  multiple D$0$-brane model with the Lorentz invariant 10D  multiple $0$-brane action of Panda and Sorokin  \cite{Panda:2003dj}.

This paper is organized as follows. In Sec. 2 we review the spinor moving frame formulation of single super-D$0$-brane (Dirichlet superparticle) and describe its irreducible kappa-symmetry. In particular, the moving frame and spinor moving frame variables used also to describe mD$0$ system are introduced there.
The mD$0$-brane action invariant under rigid (super)spacetime supersymmetry and local worldline supersymmetry ($\kappa$-symmetry) is constructed in Sec. 3. There we begin by describing 1d ${\cal N}=16$ $SU(N)$ SYM multiplet, and then make its supersymmetry local by coupling it to the composite supergravity induced on the mD$0$ worldline
(this is to say on the worldline of the center of mass of  mD$0$ system) and by inclusion of a single D$0$-brane action ('center of mass brane' action) into the complete action of the interacting system.
In Sec. 4 we compare our result with the action of multiple $0$-brane system proposed by Panda and Sorokin  \cite{Panda:2003dj} and argue that our action is better candidate for the description of multiple D$0$-brane system. We conclude in Sec. 5.

The Appendices are devoted to the problem of dimensional reduction of the 11D mM$0$ action. Although this does not give a desired result, {\it i.e.} does not reproduce the mD$0$ action, its discussion may give useful suggestion for further thinking.
In Appendix A we describe the spinor moving frame action for single M-wave (M$0$-brane) and show how its dimensional reduction reproduces the action for single 10D D$0$-brane. The dimensional reduction of spinor moving frame variables is discussed in Appendix B. Appendix C describes the  dimensional reduction of 11D mM$0$  action down to D=10. The mM$0$ action and its local worldline supersymmetry are presented  in Appendix C.1. Its dimensional reduction and an apparent difference of the result of this from mD$0$ action of Sec. 3 are discussed in Appendix C.2.

\section{D$0$-brane in moving frame formulation}

Let us denote the coordinates of flat 10D type IIA superspace $\Sigma^{(10|32)}$ by $(x^a, \theta^{\alpha 1}, \theta_\alpha^2)$ and its superveilbein by
\begin{eqnarray}\label{Ea:=}
E^a= dx^a - i d\theta^{\alpha 1} \sigma^a_{\alpha\beta}\theta^{\beta 1}-i  d\theta_\alpha^2\tilde{\sigma}_a^{\alpha\beta}d\theta_\beta^2
\; ,    \qquad E^{\alpha 1}=d\theta^{\alpha 1}\; , \qquad E_\alpha^2 =d\theta_\alpha^2\; .
\end{eqnarray}
We will use the same symbols for the  pull--back of the supervielbein forms to the worldline which depend on coordinate functions $(x^a(\tau), \theta^{\alpha 1}(\tau), \theta_\alpha^2(\tau))$ of the superparticle proper time $\tau$. These are used to define parametrically the superparticle worldline as a line in the superspace,
\begin{eqnarray}\label{W1in}
{\cal W}^1 \subset  \Sigma^{(10|32)}\quad : \qquad x^a=x^a(\tau), \qquad  \theta^{\alpha 1}= \theta^{\alpha 1}(\tau), \qquad \theta_\alpha^2=\theta_\alpha^2(\tau)\; .
\end{eqnarray}
Here and below $a,b,c=0,1,...,9$ are ten-vector indices,
$\alpha, \beta,\gamma =1,...,16$ are 10D Majorana--Weyl spinor indices and
$\sigma^a_{\alpha\beta}=\sigma^a_{\beta\alpha}$ and  $\tilde{\sigma}_a^{\alpha\beta}=\tilde{\sigma}_a^{\beta\alpha}$ are 10D generalized Pauli matrices obeying
$$
\sigma^a\tilde{\sigma}{}^b + \sigma^b\tilde{\sigma}{}^a= \eta^{ab} {\bb I}_{16\times 16} , \qquad
\eta^{ab}=diag (1,\underbrace{-1,...,-1}_9\,) \; .
$$

The moving frame formulation of the 10D D$0$-brane in flat type IIA superspace is based on the action \cite{Bandos:2000tg}
\begin{eqnarray}\label{SD0}
S_{D0} &=& M
\int_{{\cal W}^1}{E}^{a} u_a^0
-iM \int_{{\cal W}^1}(d\theta^{\alpha 1} \theta_\alpha^2-\theta^{\alpha 1} d\theta_\alpha^2)
\; ,   \qquad
\end{eqnarray}
where $M$ is a constant mass parameter and $u_a^0=u_a^0(\tau)$ is an auxiliary ten-vector field of unit length,
$u_a^0u^{a0}=1$.

\subsection{Moving frame and Cartan forms}

It is convenient to consider $u_a^0$ as one of the vectors of moving frame field described by the Lorentz group valued  matrix
  \begin{eqnarray}\label{uab=D10}
 u_a^{(b)}(\tau) = \left(  u_a^{0}, \; u_a^i \right)\; \in \; SO(1,9)\; . \qquad
\end{eqnarray}
Eq. (\ref{uab=D10}) implies orthogonality and normalization conditions
 \begin{eqnarray}\label{uauai=0}
 u_a^0u^{a0}=1, \qquad u_a^0u^{ai}=0\; , \qquad u_a^iu^{aj}=- \delta^{ij}\; , \qquad
\end{eqnarray}
so that, on one hand, the variables (\ref{uab=D10}), which are called {\it moving frame variables}, are highly constrained.
On the other hand, as the space (co)tangent to a group is isomorphic to its Lie algebra,  (\ref{uab=D10}) implies that one can easily express the derivatives and variations of the  moving frame vectors $u_a^0, u_a^I$, constrained by (\ref{uauai=0}), in terms of Cartan forms of $SO(1,9)$,
 \begin{eqnarray}\label{Omi=}
 \Omega^i= u_a^0du^{ai} , \qquad  \Omega^{ij}=u_a^idu^{aj}\; . \qquad
\end{eqnarray}
Notice that the splitting of matrix in (\ref{uab=D10}) is invariant under the local $SO(9)$ rotations and
the above $\Omega^{ij}$ transforms as a connection under these.
Hence we can define $SO(9)$ covariant derivatives and find that their action on the moving frame vectors is expressed through  the covariant Cartan form $\Omega^{i}$,
 \begin{eqnarray}\label{Du0=}
 Du_a^0=du_a^0= u_a^i \Omega^i\; , \qquad Du_a^{i}=du_a^{i} + u_a^{j}\Omega^{ji}=u_a^0 \Omega^{i}\; . \qquad
\end{eqnarray}
The admissible variations of the moving frame vectors, this is to say the variations which preserve  (\ref{uauai=0}) and hence (\ref{uab=D10}), can be obtained from (\ref{Du0=}) by formal contraction with the variation symbol,
 \begin{eqnarray}\label{varu0=}
 \delta u_a^0= u_a^i i_\delta \Omega^i\; , \qquad \delta u_a^{i} =- u_a^{j}i_\delta  \Omega^{ji}+u_a^0 i_\delta \Omega^{i}\; . \qquad
\end{eqnarray}
On this way we used $i_\delta d=\delta$ \footnote{This is a particular case of the Lie derivative formula for
general coordinate variations of differential forms, $ \delta=i_\delta d+di_\delta$, written for the case of 0-forms.} and consider the contractions of the Cartan forms $i_\delta \Omega^i$ and $i_\delta  \Omega^{ji}$ as independent variations. The latter corresponds to $SO(9)\subset SO(1,9)$, which is the manifest gauge symmetry  of our construction  (acting trivially on the action (\ref{SD0})) and the former, $i_\delta \Omega^i$, is the essential variation corresponding to the coset $SO(1,9)/SO(9)$.
Our moving frame variables can be considered as a kind of (constrained) homogeneous coordinates of such a coset.

\subsection{Covariant splitting of supervielbein, action variation and kappa-symmetry}

The moving frame vectors can be used to split the bosonic supervielbein in two parts in a Lorentz invariant manner. Indeed, just contracting the supervielbein (\ref{Ea:=}) with moving frame vectors we arrive at  one invariant bosonic 1--form and one 9-vector 1--form carrying the index of the local $SO(9)$ symmetry group
 \begin{eqnarray}\label{E0:=}
  E^0= E^au_a^0\; , \qquad E^i= E^a u_a^i \; . \qquad
\end{eqnarray}
The latter does not appear in the action, but it does in the action variation.

A simple way to calculate this latter is by using the Lie derivative formula applied to the Lagrangian one-form of the action (\ref{SD0}),
\begin{eqnarray}\label{cLD0=}
{\cal L}^{D0}_{1} &=& M {E}^0
-iM (d\theta^{\alpha 1} \theta_\alpha^2-\theta^{\alpha 1} d\theta_\alpha^2)
\; ,   \qquad
\end{eqnarray}
$\delta {\cal L}^{D0}_{1}= i_\delta d{\cal L}^{D0}_{1} + d(i_\delta {\cal L}^{D0}_{1})$. The second term does not contribute to the action variation as far as we are not interested in initial conditions, and to obtain the first we have to begin by calculating a formal exterior differential of the Lagrangian form (\ref{cLD0=}). After some algebra we obtain \footnote{In our notation the exterior derivative acts from the right, e.g.
$d (E^au_a^0)= dE^a\, u_a^0 + E^a\wedge du_a^0$ where   $\wedge $ is the exterior product of differential forms. The exterior product of bosonic forms is antisymmetric, e.g.
$E^a\wedge E^b=- E^b\wedge E^a$. Here  and in (\ref{dcLD0=}) below  one should think about differential forms on target superspace or its extension, but not just on the worldline.  }
\begin{eqnarray}\label{dcLD0=}
d{\cal L}^{D0}_{1} &=& M {E}^i\wedge \Omega^i
-iM (E^{\alpha 1} + \tilde{\sigma}^{0\alpha\gamma} E_\gamma^2)\wedge  \sigma^0_{\alpha\beta}(E^{\beta 1} + \tilde{\sigma}^{0\beta\varepsilon} E_\varepsilon^2)
\; ,   \qquad
\end{eqnarray}
where\footnote{In this paper we will use the notation $\sigma^0$ and $\sigma^i$ for Lorentz covariant projections of sigma matrices,
${\sigma}^{0}_{\alpha\gamma}:= u_a^0 {\sigma}^{a}_{\alpha\gamma}$ and ${\sigma}^{i}_{\alpha\gamma}:= u_a^i {\sigma}^{a}_{\alpha\gamma}$ (see
(\ref{u0s=vv})). Lorentz non-covariant splitting will be denoted by
$(\sigma^{(0)}_{qp}, \sigma^{(i)}_{qp})=(\delta_{qp}, \gamma^{i}_{qp})$, see below.
 }
\begin{eqnarray}\label{sig0=}
{\sigma}^{0}_{\alpha\gamma}:= u_a^0 {\sigma}^{a}_{\alpha\gamma}\; , \qquad \tilde{\sigma}^{0\alpha\gamma}:= u_a^0 \tilde{\sigma}^{a\alpha\gamma}
\; .   \qquad
\end{eqnarray}
The presence of only one linear combination of the two fermionic supervielbein forms in (\ref{dcLD0=}) indicates the local fermonic $\kappa$-symmetry  of the D$0$-brane action (\ref{SD0}) (see \cite{deAzcarraga:1982dhu,deAzcarraga:1982njd,Bergshoeff:1996tu} for the kappa-symmetry of the standard D$0$-brane action). Its transformations read
\begin{eqnarray}\label{kappa=sym}
& \delta_\kappa \theta^{\alpha 1}=\kappa^\alpha\; , & \qquad  \delta_\kappa \theta^2_{\alpha}=-\sigma^0_{\alpha\beta} \kappa^\beta\; , \qquad \nonumber \\
&& \delta_\kappa x^a=  - i \theta^{\alpha 1} \sigma^a_{\alpha\beta}\delta_\kappa \theta^{\beta 1}-i  \theta_\alpha^2\tilde{\sigma}_a^{\alpha\beta}\delta_\kappa \theta_\beta^2\; , \qquad \nonumber \\ && \delta_\kappa  u_a^0=0 \quad  (i_\kappa \Omega^i\equiv i_{\delta_\kappa} \Omega^i =0) \; . \qquad
\end{eqnarray}

As it was shown in \cite{Sorokin:1989zi}, kappa symmetry is actually a local worldline supersymmetry of the superparticle models. It will be important for our discussion below that the
moving frame formulation of the superparticle actually provides us with a composite supergravity multiplet for this local supersymmetry.

Let us consider
\begin{eqnarray}\label{N16=SG}
E^0 = E^a u_a^0\; , \qquad (E^{\alpha 1} - \tilde{\sigma}^{0\alpha\gamma}E_\gamma^2)=
d\theta^{\alpha 1} - \tilde{\sigma}^{0\alpha\gamma}d\theta_\gamma^2\; .
\end{eqnarray}
Under the $\kappa$--symmetry they transform as
\begin{eqnarray}\label{kappa=SG}
\delta_\kappa E^0 = -2i (E^{\alpha 1} - \tilde{\sigma}^{0\alpha\gamma}E_\gamma^2) \sigma^0_{\alpha\beta}\kappa^{\beta} \; , \qquad \delta_\kappa (E^{\alpha 1} - \tilde{\sigma}^{0\alpha\beta}E_\beta^2)= 2d \kappa^\alpha \; , \qquad
\end{eqnarray}
which is quite similar to the transformation of graviton and gravitino one forms of  $d=1$ ${\cal N}=16$ supergravity.
However, the identification with  supergravity is hampered by that both the counterparts of gravitini and of parameter of supersymmetry carry 10D MW spinor index. Thus we will call the fermionic form in (\ref{N16=SG}) `proto-gravitino'.
To find a true counterpart of gravitini induced by the embedding of the D$0$-brane worldline in superspace, we need to introduce one more ingredient: spinor moving frame field (also called spinor Lorentz harmonic\footnote{See \cite{Galperin:1984av,Galperin:1984bu,Galperin:2001uw} for the concept of harmonic variables and harmonic superspace and \cite{Sokatchev:1985tc,Sokatchev:1987nk,Bandos:1990ji,Galperin:1991gk,Delduc:1991ir} for Lorentz harmonics (called light-cone harmonics in \cite{Sokatchev:1985tc,Sokatchev:1987nk}).}).

\subsection{Spinor moving frame and induced worldline supergravity}

$Spin(1,9)/Spin(9)$ spinor moving frame variable is $16\times 16$ $Spin(1,9)$ valued matrix
\begin{eqnarray}\label{harmV=10D0}
v_\alpha{}^q \in Spin(1,9)\;  \qquad
\end{eqnarray}
defined up to    $Spin(9)$ gauge transformations.
This is related  to the moving frame matrix (\ref{uab=D10}) by the conditions of the sigma-matrix preservation
$u^{(b)}_a \sigma^a_{\alpha\beta}= v_{\alpha}^q \sigma^{(b)}_{qp}v_{\beta}^p$. Choosing the representation with
$\sigma^{(b)}_{qp}=(\delta_{qp}, \gamma^i_{qp})$, where
$\gamma^i_{qp}=\gamma^i_{pq}$ are $d=9$ gamma matrices obeying $\gamma^{(i}\gamma^{j)}=\delta^{ij}$, we find
\begin{eqnarray}\label{u0s=vv}
\sigma^0_{\alpha\beta}:= u_a^0 \sigma^a_{\alpha\beta}=v_\alpha{}^q v_\beta{}^q \; , \qquad
\sigma^i_{\alpha\beta}:= u_a^i \sigma^a_{\alpha\beta}=v_\alpha{}^q \gamma^i_{qp}v_\beta{}^p \; . \qquad
\end{eqnarray}

The derivatives of the spinor moving frame matrix is expressed in terms of the same SO(1,9) Cartan forms (\ref{Omi=}). It is convenient to use the $Spin(9)$ covariant derivative which, when acting on spinor moving frame, is expressed in terms of the covariant Cartan form:
\begin{eqnarray}\label{Dv=vOm}
Dv_\alpha{}^q:= dv_\alpha{}^q+ \frac 1 4 \Omega^{ij} v_\alpha{}^p\gamma^{ij}_{pq}
= \frac 1 2 \gamma^i_{qp} v_\alpha{}^p\Omega^i \; . \qquad
\end{eqnarray}

We will need also the inverse spinor moving frame matrix $v_q{}^{\alpha} \in Spin(1,9)$,
\begin{eqnarray}\label{harmV-1=10D0}
v_q{}^\alpha v_\alpha{}^p=\delta_q{}^p \qquad \Leftrightarrow \qquad
 v_\alpha{}^q v_q{}^\beta= \delta_\alpha{}^\beta
\; . \qquad
\end{eqnarray}
It can be used to factorize the  matrices with upper spinor indices
\begin{eqnarray}\label{u0ts=vv}
\tilde{\sigma}^{0\alpha\beta}:= u_a^0 \tilde{\sigma}^{a\alpha\beta}=v_q{}^\alpha v_q{}^\beta \; , \qquad
\tilde{\sigma}^{i\alpha\beta}:= u_a^i \tilde{\sigma}^{a\alpha\beta}=- v_q{}^\alpha  \gamma^i_{qp} v_p{}^\beta\; . \qquad
\end{eqnarray}
One can easily check that
\begin{eqnarray}\label{vs=v-1}
v_\alpha^q \tilde{\sigma}^{0\alpha\beta}=  v_q{}^\beta\; , \qquad  {\sigma}^0_{\alpha\beta}v_q{}^\beta  = v_\alpha{}^q\; .
\end{eqnarray}

The spinor moving frame field can be used to construct the fermionic forms with the indices of
$SO(9)$  gauge group (cf. (\ref{E0:=}))
\begin{eqnarray}\label{Eq1=}
E^{q1}= E^{\alpha 1}v_\alpha^q= d\theta^{\alpha 1}\, v_\alpha{}^q\; , \qquad  E_q^2 = E_\alpha^2v_q{}^\alpha= d\theta_\alpha^2v_q{}^\alpha\; .
\end{eqnarray}
We can also define the parameter of the worldline supersymmetry ($\kappa$-symmetry) with an internal $SO(9)$ index
\begin{eqnarray}\label{eq=kapv}
\varepsilon^q:= \kappa^\alpha v_\alpha{}^q\; .
\end{eqnarray}
This can be identified with parameter of the standard ${\cal N}=16$ extended $d=1$ supersymmetry\footnote{To be precise, we have to notice that the natural R-symmetry group $SO(16)$ of such an extended supersymmetry is broken down to $SO(9)$ in our model.}.

In particular, contracting the proto-gravitino form in (\ref{N16=SG}) with spinor frame matrix we arrive at
fermionic one form
\begin{eqnarray}\label{E1-E2=}
(E^{\alpha 1} - \tilde{\sigma}^{0\alpha\gamma}E_\gamma^2)v_\alpha^q =
E^{q1}-E_q^2\;
\end{eqnarray}
which transforms as true gravitino of $d=1$ ${\cal N}=16$ supergravity under the worldline supersymmetry.
Indeed, (\ref{kappa=SG}) can be written in the following equivalent form
\begin{eqnarray}\label{susy=SG}
\delta_\varepsilon E^0 = -2i (E^{q1}-E_q^2)\varepsilon^q\; , \qquad \delta_\varepsilon  (E^{q1}-E_q^2)=2D\varepsilon^q\; , \qquad
\end{eqnarray}
where the covariant derivative is defined in (\ref{Dv=vOm}),
\begin{eqnarray}\label{Dvare=}
D\varepsilon^q:= d\varepsilon^q+ \frac 1 4 \Omega^{ij} \varepsilon{}^p\gamma^{ij}_{pq}\; . \qquad
\end{eqnarray}

Eq. (\ref{susy=SG}) has the form of  typical supersymmetry transformations of  supergravity multiplet. In our case this multiplet is composite, induced by embedding of the super-D0-brane worldline in the flat type IIA superspace. In the next section we will construct an action for  multiple D$0$-brane system by putting
 $d=1$ ${\cal N}=16$ SYM multiplet on the worldline of a single D$0$-brane and making its supersymmetry local by coupling  it to this induced supergravity.

\section{Multiple D$0$-brane action from locally supersymmetric SYM on the worldline of a D$0$-brane}

\subsection{$d=1$ ${\cal N}=16$ SYM}

The $d=1$ ${\cal N}=16$ $SU(N)$ SYM multiplet contains three types of  $N\times N$ traceless matrix fields: 1d gauge field  ${\bb A}_\tau (\tau) $, which we prefer to include in the 1-form  ${\bb A}=d\tau {\bb A}_\tau (\tau) $, nanoplet of bosonic  fields  $ {\bb X}^i(\tau)$ in vector representation of
$SO(9)$ and hexadecuplet of fermionic matrix fields $ {\Psi}_q$ in the spinor representation of $SO(9)$. In addition, we find convenient to introduce an auxiliary bosonic   matrix fields $ {\bb P}^i(\tau)$ which play the role of momenta conjugate to  $ {\bb X}^i(\tau)$ fields.

The Lagrangian one-form for  the action of $d=1$ ${\cal N}=16$ SYM can be written as (see \cite{deWit:1988wri,Banks:1996vh,Bandos:2013uoa})
\begin{eqnarray}\label{cL=SYM}
{\cal L}_1^{SYM}=d\tau L_{SYM}= tr (- {\bb P}^i\nabla {\bb X}^i + 4i\Psi_q\nabla \Psi_q)+ d\tau {\cal H}\; , \qquad
\end{eqnarray}
where
\begin{eqnarray}
\label{SYMDX=}
 \nabla {\bb  X}^i= d{\bb X}^i+ [{\bb A},    {\bb X}^i]\; , \qquad \nabla {\Psi}_q= d{\Psi}_q+ [{\bb A},   {\Psi}_q]\;
  \end{eqnarray}
are SYM covariant derivatives of the scalar and spinor fields and  ${\cal H}$ is the SYM Hamiltonian
\begin{eqnarray}
\label{HSYM=1}
{\cal H}=   {1\over 2} tr\left( {\bb P}^i {\bb P}^i \right) - {1\over 64}
tr\left[ {\bb X}^i ,{\bb X}^j \right]^2 - 2\,  tr\left({\bb X}^i\, \Psi\gamma^i {\Psi}\right) \;   \qquad
  \end{eqnarray}
  which contains the positively definite scalar potential
\begin{eqnarray}
\label{VSYM=} {\cal V} = - {1\over 64}
tr\left[ {\bb X}^i ,{\bb X}^j \right]^2 \equiv  +{1\over 64}  tr\left[ {\bb X}^i
,{\bb X}^j \right] \cdot \left[ {\bb X}^i ,{\bb X}^j \right]^\dagger  . \qquad
  \end{eqnarray}
In the  last term of (\ref{HSYM=1}), which describes the  Yukawa coupling of the bosonic and fermionic matrix fields,
  $\gamma^i_{qp}$ are the 9d Dirac matrices. They are real, symmetric, $\gamma^i_{qp}=\gamma^i_{pq}$,  and obey the
Clifford algebra \begin{eqnarray}\label{gigj+=} \gamma^i\gamma^j + \gamma^j \gamma^i=
2\delta^{ij} I_{16\times 16}\; , \qquad
\end{eqnarray}
as well as the following identities
\begin{eqnarray}\label{gi=id1}
&& \gamma^{i}_{q(p_1}\gamma^{i}_{p_2p_3) }= \delta_{q(p_1}\delta_{p_2p_3) }\; , \qquad
 \gamma^{ij}_{q(q^\prime }\gamma^{i}_{p^\prime)p }+
\gamma^{ij}_{p(q^\prime }\gamma^{i}_{p^\prime)q } = \gamma^{j}_{q^\prime
p^\prime}\delta_{qp}-\delta_{q^\prime p^\prime}\gamma^{j}_{qp} \; .
\end{eqnarray}

The action $\propto\int_{W^1} {\cal L}^{SYM}_1$ is invariant under the rigid $d=1$ ${\cal N}=16$ supersymmetry transformations with constant fermionic parameter $\varepsilon^q$
\begin{eqnarray}
\label{SYMsusy-X} \delta_\varepsilon {\bb X}^i   = 4i \varepsilon^q (\gamma^i  \Psi)_q \; , \qquad
\delta_\varepsilon {\bb P}^i   = [\varepsilon^{q} (\gamma^{ij}  \Psi)_q,  {\bb X}^j]\; ,\qquad \\
\label{SYMsusy-Psi} \delta_\varepsilon \Psi_q =  {1\over 2} \varepsilon^{p} \gamma^i_{pq}  {\bb
P}^i-  {i\over 16} \varepsilon^{p} \gamma^{ij}_{pq}  [{\bb X}^i, {\bb X}^j]\; ,\qquad \\
\label{SYMsusy-A}
 \delta_\varepsilon {\bb A}  = - d\tau   \varepsilon^{q}  \Psi_q
 \; .  \qquad
\end{eqnarray}

Notice that  supersymmetry acts on the SYM Hamiltonian (\ref{HSYM=1}) by
\begin{eqnarray}
\label{cH=susy}
\delta_\varepsilon {\cal H} =  \varepsilon^q tr (\Psi_q{\bb G})\; ,
  \qquad
\end{eqnarray}
where
\begin{eqnarray}
\label{G=Gauss}
 {\bb G}=
 [{\bb P}^i, {\bb X}^i] - 4i \{ \Psi_q, \Psi_q\}
  \qquad
\end{eqnarray}
is the Gauss law constraint which appears as equation of motion for the 1d gauge field of ${\cal N}=16$ SYM model.
In the action variation (\ref{cH=susy}) is compensated by the nontrivial supersymmetry transformation (\ref{SYMsusy-A}) of the 1d gauge field.
The Gauss law is supersymmetric invariant,
\begin{eqnarray}
\label{G=susy}
\delta_\varepsilon {\bb G}=0\; .
  \qquad
\end{eqnarray}

\subsection{From SYM to mD$0$ brane action}

As we have already announced, the multiple D$0$-brane (mD$0$) action can be obtained on the way of putting the maximally supersymmetric  $SU(N)$ SYM multiplet on the worldline of a single D$0$-brane (center of mass brane of the mD$0$ system) and coupling  it to supergravity induced by embedding of this worldline into the tangent superspace. Let us describe the procedure in detail.

First of all, let us consider the variation of the SYM Lagrangian form under supersymmetry (\ref{SYMsusy-X})--(\ref{SYMsusy-A}) with local fermionic parameter $\varepsilon^q$. This gives\footnote{Notice that to establish supersymmetry invariance of the action, one has to perform integration by parts. This fact has also to be taken into account carefully to establish the correct coefficients for $\propto d\varepsilon^q$ terms. We do not write explicitly  the corresponding total derivative terms in our expression for $\delta {\cal L}^{SYM}_1$. }
\begin{eqnarray}
\label{susy=de}
\delta {\cal L}^{SYM}_1 = -4i d\varepsilon^q tr (\gamma^i_{qp}\Psi_p{\bb P}^i)- \frac 1 2  d\varepsilon^q tr (\gamma^{ij}_{qp}\Psi_p [{\bb X}^i, {\bb X}^j ])\; .
  \qquad
\end{eqnarray}
 According to the first Noether theorem this implies that
\begin{eqnarray}
\label{S-current}
{\cal S}_q= 2 tr (\gamma^i_{qp}\Psi_p{\bb P}^i)- \frac i 4  tr (\gamma^{ij}_{qp}\Psi_p [{\bb X}^i, {\bb X}^j ])\;
  \qquad
\end{eqnarray}
is the supercurrent for the rigid supersymmetry of the 1d ${\cal N}=16$ SYM.

To construct the action invariant under local supersymmetry, following the Noether procedure, we include into the Lagrangian form  the new term given by the product of gravitino and supercurrent,
\begin{eqnarray}
\label{cL2=ES}
{\cal L}_1^2 = i (E^{q1}-E_q^2) {\cal S}_q\; .
  \qquad
\end{eqnarray}

At this stage we notice that the induced gravitino transformations (\ref{susy=SG}) include covariant derivative $D$ of the supersymmetry parameter,
(\ref{Dvare=}),  rather then the usual derivative. The Lagrangian form which will provide the transformations of the form like in  (\ref{susy=de}) but with covariant derivatives
(\ref{Dv=vOm}) will be obtained by replacing in ${\cal L}_1^{SYM}$ $\nabla$ by $D$ including also the $SO(9)$ connection as in (\ref{Dv=vOm}),
\begin{eqnarray}
\label{DXi=10D} D{\bb X}^i  &:=& d{\bb X}^i   - \Omega^{ij} {\bb
X}^j+ [{\bb A},    {\bb X}^i] \; , \qquad \\ \label{DPsi:=10D} D\Psi_q  &:=& d\Psi_q
 -{1\over 4} \Omega^{ij} \gamma^{ij}_{qp} {\Psi}_p+ [{\bb A},
 \Psi_q ] \; . \qquad
  \end{eqnarray}

But this is still not the end of story. Notice that the supersymmetry transformation of the supercurrent is
\begin{eqnarray}
\label{S-cur=susy}
\delta_\varepsilon {\cal S}_q= 2 \varepsilon^q {\cal H} -\frac i 2  \gamma^i_{qp}\varepsilon^p tr ({\bb G}{\bb X}^i)\; ,
  \qquad
\end{eqnarray}
where ${\cal H}$ is the SYM Hamiltonian (\ref{HSYM=1}) and ${\bb G}$ is the Gauss law constraint  (\ref{G=Gauss}).
The corresponding contributions to the  variation of ${\cal L}_1^2$ (\ref{cL2=ES}) can be compensated if we replace in  ${\cal L}_1^{SYM}$ (\ref{cL=SYM})
$d\tau $ by $E^0$ of (\ref{N16=SG}), thus providing the coupling of SYM sector to 1d induced 'graviton',
 and by modifying the transformation rule of the 1d gauge field. This latter is achieved by changing
 $d\tau \mapsto E^0$ in (\ref{SYMsusy-A}) and by adding the term $1/2 ({E}^{q1}-E_q^2)\gamma^i_{qp}
 \epsilon^{p}\;    {\bb X}^i$ to this transformation rule (see below).
 Thus the multiple D$0$-brane action should contain the following modification of the SYM Lagrangian form
\begin{eqnarray}
\label{cL1=LSYM}
{\cal L}_1^1 = {\cal L}_1^{SYM}\vert_{\nabla\mapsto D\; , \; d\tau \mapsto E^0}\; .
  \qquad
\end{eqnarray}

Resuming, the locally supersymmetric ($\kappa$--symmetric) invariant action for multiple  D$0$-brane system is the integral of Lagrangian 1--form given by the sum of (\ref{cLD0=}),  (\ref{cL1=LSYM}) and (\ref{cL2=ES}),
\begin{eqnarray}
\label{cL1mD0=}
{\cal L}_1^{mD0}= {\cal L}_1^{D0} + k{\cal L}_1^1+ k{\cal L}_1^2\; .
 \qquad
\end{eqnarray}
Here we have introduced a constant $k$ of dimension \footnote{Notice that
$[{\bb X}^i]=[M]$, $[{\bb P}^i]=[M]^2$, $[\Psi_q]= [M]^{3/2}$, which reflects the SYM origin of these matrix fields.} $[M]^{-3}$. The presence of  the Lagrangian form of the single D$0$-brane action (\ref{SD0}), ${\cal L}_1^{D0}$, is necessary in (\ref{cL1mD0=}) to make nontrivial the center of mass dynamics described by the equations for the coordinate functions (\ref{W1in}).

\subsection{Multiple D$0$ brane action and its local worldline supersymmetry}

For the reader convenience we write the complete form of the above described  multiple D$0$-brane  action explicitly:
\begin{eqnarray}\label{SmD0=}
S_{mD0} = \int_{{\cal W}^1} {\cal L}_1^{mD0}&=& M
\int_{{\cal W}^1}\left( {E}^0
-i(d\theta^{\alpha 1} \theta_\alpha^2-\theta^{\alpha 1} d\theta_\alpha^2)\right)+
\nonumber \\ && + k\int_{{\cal W}^1}\left(tr (- {\bb P}^iD{\bb X}^i + 4i\Psi_qD \Psi_q)+ E^0 {\cal H}\right)+
\nonumber \\ && +
2i k\int_{{\cal W}^1} (E^{q1}-E_q^2)  tr \left(\gamma^i_{qp}\Psi_p{\bb P}^i- \frac i 8  \gamma^{ij}_{qp}\Psi_p [{\bb X}^i, {\bb X}^j ]\right)\; .
  \qquad
\end{eqnarray}
In it ${E}^0$ is given by the contraction (\ref{E0:=}) of the pull--back of supervielbein (\ref{Ea:=}) with moving frame vector
(see (\ref{uab=D10})), $E^{q1}$ and $E_q^2$ are given by contractions (\ref{Eq1=}) of the pull--back of the fermionic supervielbein forms with the spinor moving frame matrices (\ref{harmV=10D0}) and
(\ref{harmV-1=10D0}), the covariant derivatives $D$ are defined in (\ref{DXi=10D}) and (\ref{DPsi:=10D}) with the use of 1d gauge field
${\bb A}$ and Cartan forms (\ref{Omi=}), and ${\cal H}$ is the SYM Hamiltonian defined in (\ref{HSYM=1}).


The action (\ref{SmD0=}) is invariant under the following local worldline supersymmetry transformations
\begin{eqnarray}
\label{susy-th} \delta_\varepsilon {\theta}^{\alpha 1} &=& \varepsilon^{q} (\tau)
v_q{}^{\alpha} \; , \qquad \delta_\varepsilon {\theta}_{\alpha}^{2} = -\varepsilon^{q} (\tau)
v_{\alpha}{}^q \; , \qquad
 \\
\label{susy-x} \delta_\varepsilon {x}^a &=& - i {\theta}^1 \sigma^a\delta_\varepsilon
{\theta}^1 - i {\theta}^2 \tilde{\sigma}{}^a\delta_\varepsilon
{\theta}^2 \; , \qquad
 \\ \label{susy-v}
 \delta_\varepsilon  v_q{}^{\alpha}&=&0
  \;   \Rightarrow  \quad  \delta_\varepsilon
  u_a^{0}= \delta_\varepsilon u_a^{i}=0\;
 ,  \qquad \\ \nonumber && {}
\\
\label{susy-X}  \delta_\varepsilon {\bb X}^i   &=& 4i \varepsilon\gamma^i  \Psi \; , \qquad
\delta_\varepsilon {\bb P}^i   = [(\varepsilon \gamma^{ij}  \Psi),  {\bb X}^j]\; ,\qquad \\
\label{susy-Psi}  \delta_\varepsilon \Psi_q &=&  {1\over 2} (\varepsilon \gamma^i)_q  {\bb
P}^i-  {i\over 16} (\varepsilon \gamma^{ij})_q  [{\bb X}^i, {\bb X}^j]\; ,\qquad \\
\label{susy-A}
 \delta_\varepsilon {\bb A} & =& -  {E}^{0} \varepsilon^{q}  \Psi_q  + \frac 1 2
 (E^{q1}-E_q^2)\gamma^i_{qp}
 \varepsilon^p\;    {\bb X}^i
 \; .  \qquad
\end{eqnarray}
The local supersymmetry transformations of the center of mass variables (coordinate functions and (spinor) moving frame variables) (\ref{susy-th})--(\ref{susy-v})
 coincide with the D$0$--brane $\kappa$--symmetry transformations (\ref{kappa=sym}) up to redefinition of the supersymmetry parameter ($\kappa^\alpha = \varepsilon^qv_q^\alpha$).
The transformations of the physical fields of 1d ${\cal N}=16$ SYM, (\ref{susy-X}) and (\ref{susy-Psi}), have the same form as in the case of rigid supersymmetry.

By construction, (\ref{SmD0=}) is also invariant under the rigid spacetime supersymmetry, which acts nontrivially on the center of mass variables only,
\begin{eqnarray}
\label{IIAsusy-xth} \delta_\varepsilon {x}^a &=& i {\theta}^1 \sigma^a\epsilon
^1 + i {\theta}^2 \tilde{\sigma}{}^a\epsilon
^2 \; , \qquad\delta_\epsilon {\theta}^{\alpha 1} = \epsilon^{\alpha 1}
 \; , \qquad \delta_\epsilon {\theta}_{\alpha}^{2} = \epsilon_{\alpha}{}^2 \; , \qquad
 \\ \label{IIAsusy-v}
 && \delta_\epsilon  v_q{}^{\alpha}=0
  \;   \Rightarrow  \quad  \delta_\epsilon
  u_a^{0}= \delta_\epsilon u_a^{i}=0\;
 ,  \qquad
\\
\label{IIAsusy-XPA}  \delta_\epsilon {\bb X}^i   &=& 0 \; , \qquad \delta_\epsilon \Psi_q = 0  \; ,\qquad
\delta_\epsilon {\bb P}^i   = 0\; ,\qquad
 \delta_\varepsilon {\bb A} =0
 \; .  \qquad
\end{eqnarray}

It is tempting to try to obtain our multiple D$0$-brane action (\ref{SmD0=}) by dimensional reduction of the 11D multiple M$0$-brane action of \cite{Bandos:2012jz}. In Appendix C we discuss such a dimensional reduction and point out a problem which appears on this way.
In the next section we discuss the differences of our multiple D$0$-brane action with very interesting multiple $0$-brane system of \cite{Panda:2003dj} and argue in favour of that rather  our (\ref{SmD0=}) is the representative of the family of mD$p$-brane actions.

\section{Differences with Panda-Sorokin multiple $0$-brane action}

In our notation the (Lorentz-covariant) action by Panda and Sorokin \cite{Panda:2003dj} reads\footnote{In  \cite{Panda:2003dj} also the actions breaking Lorentz covariance explicitly, in the same manner as it was broken in Myers action \cite{Myers:1999ps}, were considered. For our purposes it is sufficient to consider Lorentz covariant representatives of this family of the actions.}
\begin{eqnarray}
\label{S=PS}
 S_{PS} &=&\int_{{\cal W}^1} p_aE^a - \frac 1 2 \int_{{\cal W}^1} d\tau e(\tau) \left(p_ap^a -({\cal M}({\bb X}, {\bb P},\Psi))^2 \right)- \quad\nonumber \\ && - i
  \int_{{\cal W}^1} {\cal M}({\bb X}, {\bb P},\Psi) (d{\theta}^{\alpha 1}\,{\theta}_{\alpha}^{2} - {\theta}^{\alpha 1}d{\theta}_{\alpha}^{2} )
 +  \int_{{\cal W}^1} Tr\left(- {\bb P}^i d{\bb X}^i + 4i\Psi_qd\Psi_q \right) \; .  \qquad
\end{eqnarray}
In it $p_a=p_a(\tau)$ is the auxiliary 1d field having the meaning of ten--momentum conjugate to the center of mass coordinate function $x^a(\tau)$, $e(\tau)$ is an auxiliary einbein field,
and
${\cal M}({\bb X}, {\bb P},\Psi)$ is an arbitrary function of the $su(N)$ valued matrix fields, bosonic nanoplets  ${\bb X}^i$ and ${\bb P}^i $ and fermionic hexadecuplet  $\Psi_q$.

Notice that the 1d gauge field is absent in this action which thus posesses only rigid $SU(N)$ symmetry
(see recent \cite{Maldacena:2018vsr,Berkowitz:2018qhn} for discussing the differences of the standard and ungauged Matrix models of
\cite{deWit:1988wri,Banks:1996vh} and \cite{Berenstein:2002jq}). But this is not the only difference of (\ref{S=PS}) with our multiple D$0$-brane action.
Probably the most important is that the $\kappa$--symmetry transformation leaving invariant the action  (\ref{S=PS}),
\begin{eqnarray}
\label{susy-thPS} \delta_\kappa {\theta}^{\alpha 1} &=& \kappa^{\alpha} \; , \qquad \delta_\kappa {\theta}_{\alpha}^{2} = - \frac 1 {\cal M}  p_a \sigma^a_{\alpha\beta} \kappa^{\beta}  \; , \qquad
 \\
\label{susy-xPS} \delta_\kappa {x}^a &=& - i {\theta}^1 \sigma^a\delta_\kappa
{\theta}^1 - i {\theta}^2 \tilde{\sigma}{}^a\delta_\kappa
{\theta}^2 \; , \qquad
\\
\label{susy-XPS}  \delta_\kappa {\bb X}^i   &=& -i \kappa^\alpha \left(\theta_\alpha^2- \frac 1 {\cal M}  p_a \sigma^a_{\alpha\beta} \theta^{\beta 1}\right) \frac { \partial {\cal M}} {\partial {\bb P}^i}   \; , \qquad
\\
\label{susy-PPS}   \delta_\kappa  {\bb P}^i   &=& i \kappa^\alpha \left(\theta_\alpha^2- \frac 1 {\cal M}  p_a \sigma^a_{\alpha\beta} \theta^{\beta 1}\right) \frac { \partial {\cal M}} {\partial {\bb X}^i}   \; , \qquad \\
\label{susy-PsiPS}  \delta_\kappa \Psi_q &=&  -{1\over 8}  \kappa^\alpha \left(\theta_\alpha^2- \frac 1 {\cal M}  p_a \sigma^a_{\alpha\beta} \theta^{\beta 1}\right) \frac { \partial {\cal M}} {\partial  \Psi_q}
 \; , \qquad
\end{eqnarray}
transform all the matrix fields by the expression proportional to the linear combination $\left(\theta_\alpha^2- \frac 1 {\cal M}  p_a \sigma^a_{\alpha\beta} \theta^{\beta 1}\right)$ of the 'center of mass' fermionic variables $(\theta^{\alpha 1}, \theta_\alpha^2)$. In contrast, the  $\kappa$-symmetry transformations  (\ref{susy-th})--(\ref{susy-Psi}) leaving invariant our action (\ref{SmD0=}) coincide with  the local version of the SYM supersymmetry transformations.
Just this property is expected from the $\kappa$--symmetry of the multiple D$0$-brane action, the low energy limit of which should be given (in its gauge fixed version) by
$U(N)$ SYM model in which the $U(1)$ sector is not mixed by the 1d supersymmetry  with the $SU(N)$ sector.

Neither spacetime supersymmetry is expected to mix the $U(1)$ and $SU(N)$ sectors in the low energy limit of the multiple D$p$-brane action.
Such a mixture is however produced by the spacetime supersymmetry transformations leaving invariant the
Panda-Sorokin action (\ref{S=PS}):
\begin{eqnarray}
\label{Susy-thPS} \delta_\epsilon {\theta}^{\alpha 1} &=& \epsilon^{\alpha 1} \; , \qquad \delta_\epsilon {\theta}_{\alpha}^{2} = \epsilon_{\alpha}^{2} \; , \qquad
 \\
\label{Susy-xPS} \delta_\epsilon {x}^a &=& i {\theta}^1 \sigma^a\epsilon^1 + i {\theta}^2 \tilde{\sigma}{}^a\epsilon^2 \; , \qquad
\\
\label{Susy-XPS}  \delta_\epsilon {\bb X}^i   &=& -i \left(\epsilon^{\alpha 1}\theta_\alpha^2- \theta^{\alpha 1}\epsilon_\alpha^2\right) \frac { \partial {\cal M}} {\partial {\bb P}^i}   \; , \qquad
\\
\label{Susy-PPS}   \delta_\epsilon  {\bb P}^i   &=& i \left(\epsilon^{\alpha 1}\theta_\alpha^2- \theta^{\alpha 1}\epsilon_\alpha^2\right) \frac { \partial {\cal M}} {\partial {\bb X}^i}   \; , \qquad \\
\label{Susy-PsiPS}  \delta_\epsilon \Psi_q &=&  -{1\over 8} \left(\epsilon^{\alpha 1}\theta_\alpha^2- \theta^{\alpha 1}\epsilon_\alpha^2\right) \frac { \partial {\cal M}} {\partial  \Psi_q}
 \; . \qquad
\end{eqnarray}

To resume, as far as a candidate for a complete description of multiple D$0$-brane system is searched for,
an advantage of our model (\ref{SmD0=}) over the Panda-Sorokin multiple $0$-brane action (\ref{S=PS})  is that   the supersymmetry and $\kappa$-symmetry leaving invariant (\ref{SmD0=}) have the properties expected from the well known very low energy limit of multiple D$0$-brane action. Namely, the $\kappa$-symmetry of  (\ref{SmD0=}) acts on the internal sector described by traceless matrix fields as the supersymmetry of maximal 1d SU(N) SYM  which is made local by coupling to 1d supergravity induced by embedding of the center of mass worldline into the target superspace (see (\ref{susy-th})--(\ref{susy-A})). The spacetime supersymmetry of  (\ref{SmD0=}) acts on the center of mass variables only (see (\ref{IIAsusy-xth})--(\ref{IIAsusy-XPA})).
  These properties are  in contrast to the ones of the Panda-Sorokin model (\ref{S=PS}) in which both supersymmetry and $\kappa$--symmetry transformations of matrix variables
   involve the center of energy fermionic variables, the property which is not observed in the (very) low energy limit given by just U(N) SYM action.

The above observations  allow us to conclude that the action  (\ref{SmD0=})
is a better candidate for the description of  multiple D$0$-brane system. The meaning of the Panda-Sorokin action  (\ref{S=PS}) and its role in String theory is an interesting question to be thought about.


\section{Conclusion}

In this paper we have constructed the complete supersymmetric action (\ref{SmD0=}) for the system of N nearly coincident D$0$-branes (mD$0$ system) in flat ten dimensional type IIA superspace. The set of its dynamical variables can be split into two sets: the center of mass variables, which are the same as used for the description of single
D$0$-brane, and the internal variables which are described by the matrix fields forming the multiplet of
${\cal N}=16$ supersymmetric  $d=1$ SU(N) Yang-Mills theory (SYM).
The mD$0$ action is invariant under rigid spacetime supersymmetry and  local worldline supersymmetry. The rigid supersymmetry acts on the center of mass variables only. The local worldline supersymemtry acts on all the fields. On the center of mass fields it acts exactly like the kappa--symmetry of single D$0$-brane action, while on the physical fields of the internal, SYM sector it acts as local version of the SYM supersymmetry.

These set of properties is exactly what is expected from the action of multiple D$0$-brane system. In particular, they are in consonance with the statement that at the very low energy limit and upon gauge fixing of local supersymmetry, our functional reduces to the $U(N)$ SYM action, as it should be with mD$0$ action according to \cite{Witten:1995im}.

These properties are not shown by multiple 0-brane action (\ref{S=PS}) proposed in \cite{Panda:2003dj}. We discuss it in comparison with our action and noticed some essential differences. In particular, the local worldline supersymmetry transformations of all the $su(N)$ valued matrix fields of  (\ref{S=PS}) involve essentially the fermionic center of mass coordinates: are proportional to these. In searching for interrelation of the models,
one might have a hope that the difference comes from the fact that the action  of \cite{Panda:2003dj} does not contain 1d gauge field and thus should be literally compared rather with the gauge fixed version of our action (\ref{SmD0=}). Indeed, taking a look on the local supersymmetry transformations of the $SU(N)$ gauge field ${\bb A}$, (\ref{susy-A}), one confirms
that in the gauge ${\bb A}=0$ the terms  with
(derivatives of the) fermionic center of mass coordinates do appear in the transformation rules of the physical matrix fields of our model (generated by compensated gauge transformations designed to preserve the gauge ${\bb A}=0$). However, besides these new terms are clearly different from the ones characterizing the kappa-symmetry of the Panda-Sorokin action,
the initial terms in (\ref{susy-X})--(\ref{susy-Psi}) are still present in the ${\bb A}=0$ gauge and provide the terms independent on center of mass fermionic coordinate which are desired for correspondence with U(N) SYM supersymmetry at very low energy. Furthermore, even in the gauge  ${\bb A}=0$ the rigid spacetime supersymmetry of the action (\ref{SmD0=}) acts on the center of mass variables only, while the  rigid supersymmetry of Panda-Sorokin action (\ref{S=PS}) acts on the $su(N)$ valued field and also mix them with the center of mass degrees of freedom.

This allows us to conclude that our action (\ref{SmD0=}) is better candidate for the complete supersymmetric description of multiple D$0$-brane system than (\ref{S=PS}). The meaning of Panda-Sorokin action (\ref{S=PS}) in M-theoretical perspective is an interesting subject for future thinking.

It is not difficult to observe that the bosonic limit of our action does not coincide with the $p=0$ representative of the family of Dielectric brane actions by Myers \cite{Myers:1999ps}. The advantage of our action is its manifest Lorentz invariance and also that it includes fermions and possess supersymmetry and $\kappa$--symmetry. The supersymmetric and $\kappa$--symmetric version of Lorentz noninvariant Myers action was searched for during many years and  is still not known. On the other hand, the widely appreciated advantage of the family of Dielectric brane actions
\cite{Myers:1999ps} is that, identifying these with mD$p$--branes, one can explicitly relate  mD$p$-brane and nD$(p\pm 1)$-brane actions by T-duality transformations. To check whether our Lorentz covariant and doubly supersymmetric construction can provide similar result, we need to construct in our approach, in addition to mD$0$-brane action, at least the action for mD$1$-brane (multiple Dirichlet strings). The search for such an action, as well as for mD$p$ action with $p>1$, is presently on the way.

To conclude, let us point out one more puzzle.
As the dimensional reduction of single M$0$-brane (M-wave) action produces the
action of single 10D D$0$--brane \cite{Bergshoeff:1996tu}, it was natural to expect that  the mD$0$ action can be reproduced by dimensional reduction of an action for multiple M-wave (mM$0$-system). Such an mM$0$ action was constructed in  \cite{Bandos:2012jz} but, as we show in Appendix C, its dimensional reduction does not reproduce a simple action for mD$0$-brane with expected properties; in particular we have not succeed  in reproducing our (\ref{SmD0=}) action by such a dimensional reduction. The resolution of this issue or a deeper understanding of the nature of the problem is an important subject for future study.


\acknowledgments{ This work  was supported in part by the Spanish Ministry of Economy, Industry and Competitiveness  grant FPA 2015-66793-P, partially financed with FEDER/ERDF (European Regional Development Fund of the European
Union), by the Basque Government Grant IT-979-16, and the Basque Country University program UFI 11/55.
The author is thankful to Dima Sorokin for useful discussions and suggestions and to the Theoretical Department of CERN for hospitality and support of his visit on one of the final stages of this work.
}

\appendix
\setcounter{equation}0
\def\theequation{A.\arabic{equation}}

\section{ D$0$-brane action from dimensional reduction of M$0$ action in moving frame formulation}

In this appendix we describe  how the moving frame action of a single D$0$-brane can be obtained by dimensional reduction of the spinor moving frame action for M$0$-brane (M-wave). For the standard Brink-Schwarz--like formulation such a dimensional reduction of M-wave action was discussed in
\cite{Bergshoeff:1996tu}. The presence of moving frame brings some additional specific problems for dimensional reduction. However,  its use is necessary to discuss the dimensional reduction of a multiple M-wave system as for today the only known complete mM$0$ action \cite{Bandos:2012jz} is formulated within the spinor moving frame approach.

\subsection{Moving frame action for 11D M$0$-brane}

M$0$ brane action in moving frame formulation reads \cite{Bandos:2007mi,Bandos:2007wm}
\begin{eqnarray}\label{SM0=}
S_{M0} &=& \int_{{\cal W}^1}\rho^\# \underline{E}^{=}=  \int_{{\cal W}^1}\rho^\# \underline{E}^{\underline{a}}U_{\underline{a}}^{=}\; . \qquad
\end{eqnarray}
Here and below $\underline{E}^{\underline{a}}$ and $\underline{E}^{\underline{\alpha}}$ are pull--backs of the supervielbein forms of 11D flat superspace
\begin{eqnarray}\label{Eua=}\underline{E}^{\underline{a}}= dX^{\underline{a}}- id\Theta \Gamma^{\underline{a}}\Theta\; , \qquad
\underline{E}^{\underline{\alpha}}= d\Theta^{\underline{\alpha}}\; , \qquad \underline{a}=0,1,...,9,10\; , \qquad \underline{\alpha}=1,...,16\; ,
\end{eqnarray}
$X^{\underline{a}}=X^{\underline{a}}(\tau)$ and $\Theta^{\underline{\alpha}}=\Theta^{\underline{\alpha}}(\tau)$ are coordinate functions describing parametrically the embedding of worldline ${\cal W}^1$ in 11D superspace
and $U_{\underline{a}}^{=}=U_{\underline{a}}^{=}(\tau)$ is a light-like vector.
It is convenient to consider it as difference of two columns of the
$SO(1,10)$ valued moving frame matrix field
\begin{eqnarray}\label{Uab=in11}
 U_{\underline{a}}^{({\underline{b}})}(\tau) = \left( {1\over 2}\left( U_{\underline{a}}^{=}+U_{\underline{a}}^{\#}
 \right), \; U_{\underline{a}}^i \, , {1\over 2}\left( U_{\underline{a}}^{\#}-U_{\underline{a}}^{=}
 \right)\right)\; \in \; SO(1,10)\; . \qquad
\end{eqnarray}
As in the case of 10D D$0$-brane, the moving frame can be used to split the pull--back of the bosonic supervielbein form (\ref{Eua=}) in a Lorentz invariant manner. In our case this will be the splitting  into two singlets and one nanoplet of the $SO(1,1)\otimes SO(9)$ gauge symmetry group,
\begin{eqnarray}\label{E--=Eu--}
\underline{E}^{=}:= \underline{E}^{\underline{a}} U_{\underline{a}}^{=} \; , \qquad \underline{E}^{\#}:= \underline{E}^{\underline{a}}U_{\underline{a}}^{\#}\; , \qquad
 \underline{E}^i:= \underline{E}^{\underline{a}} U_{\underline{a}}^i \, . \qquad
\end{eqnarray}

\subsection{Dimensional reduction of the moving frame action. D$0$ from M$0$  }

To perform a dimensional reduction of the M$0$ action (\ref{SM0=}) down to 10 dimension we should relate the
11D moving frame matrix (\ref{Uab=in11}) to its 10D cousin (\ref{uab=D10}),
\begin{eqnarray}\label{uab=in10}
 u_a^{(b)}(\tau) = \left(  u_a^{0}, \; u_a^i \right)\; \in \; SO(1,9)\; , \qquad a=0,1,...,9\; , \qquad i=1,...,9\;.
\end{eqnarray}
To this end we use $SO(1,10)$ valued matrix
$L_{\underline{a}}{}^{\underline{b}}= (L_{\underline{a}}{}^b , L_{\underline{a}}{}^*)$ representing the coset
$SO(1,10)/SO(1,9)$. The generic relation reads
\begin{eqnarray}\label{U=Lu}
 U_{\underline{a}}^{({\underline{b}})} = L_{\underline{a}}{}^{\underline{c}} \left(\begin{matrix}
 u_c^{(b)} & 0 \cr 0 & 1
 \end{matrix}\right)=   \left(\begin{matrix}
L_{{a}}{}^{{c}}  u_c^{(b)} & L_{{a}}{}^{*} \cr L_{*}{}^{{c}}  u_c^{(b)} & L_{*}{}^{*}
 \end{matrix}\right)\;  \qquad
\end{eqnarray}
and implies
\begin{eqnarray}\label{U--=Lu-L}
  U_{\underline{a}}^{=}= L_{\underline{a}}{}^{{c}}  u_c^{0} - L_{\underline{a}}{}^{*}\; , \qquad
\\ \label{U++=Lu+L}
  U_{\underline{a}}^{\#}= L_{\underline{a}}{}^{{c}}  u_c^{0} + L_{\underline{a}}{}^{*}\; , \qquad
   \\ \label{UI=LuI}
   U_{\underline{a}}^i =  L_{\underline{a}}{}^{{c}}  u_c^{i}\; . \qquad
\end{eqnarray}

To perform the dimensional reduction, let us  firstly write M$0$ action (\ref{SM0=}) in  terms of 10D moving frame variables and L-matrix,
\begin{eqnarray}\label{SM0=10+1}
S_{M0} &=&
\int_{{\cal W}^1}\rho^\# {E}^{\underline{a}}L_{\underline{a}}{}^{b} u_b^0 -
\int_{{\cal W}^1}\rho^\# {E}^{\underline{a}}L_{\underline{a}}{}^{*}
\; .    \qquad
\end{eqnarray}
Secondly, let us consider $L$-matrix to be constant, so that
\begin{eqnarray}\label{E*=}
\underline{E}^*= \underline{E}^{\underline{a}}L_{\underline{a}}{}^{*} =dX^* - i d\Theta \Gamma^* \Theta  \; , \qquad \Gamma^*:= \Gamma^{\underline{a}}L_{\underline{a}}{}^{*}\; , \\
\label{Eb=ELb=}
\underline{E}^b:=\underline{E}^{\underline{a}}L_{\underline{a}}{}^{b}= dX^b - i d\Theta \Gamma^b \Theta  \; , \qquad \Gamma^b:= \Gamma^{\underline{a}}L_{\underline{a}}{}^{b}\; ,
\end{eqnarray}
and the action (\ref{SM0=10+1}) becomes
\begin{eqnarray}\label{SM0=10+1=}
S_{M0}\vert_{dL=0} &=&
\int_{{\cal W}^1}\rho^\# \underline{E}^{b} u_b^0 -
\int_{{\cal W}^1}\rho^\# (dX^* - i d\Theta \Gamma^* \Theta)
\; .    \qquad
\end{eqnarray}

Now, if we   consider $\rho^\# $ to be a constant,
\begin{eqnarray}\label{r++=M}\rho^\# =M=const\; , \qquad
\end{eqnarray}
then $X^*$ coordinate drops from  the action which reads
\begin{eqnarray}\label{SM0=SD0}
S_{M0}\vert_{dL=0=d \rho^\#} &=& M
\int_{{\cal W}^1}\underline{E}^{b} u_b^0
+iM \int_{{\cal W}^1}d\Theta \Gamma^* \Theta
\;     \qquad
\end{eqnarray}
and can be recognized as D$0$-brane action (\ref{SD0}).
To reach the literal coincidence, we have to   split the 10D Majorana spinor fermionic coordinates in two Majorana-Weyls spinors,
\begin{eqnarray}
\label{Th=th12}
\Theta^{\underline{\alpha}}=\left( \begin{matrix}\theta^{{\alpha}1} \cr \theta_{{\alpha}}^2 \end{matrix}\right)
\end{eqnarray}
and use the gamma matrix representation with
\begin{eqnarray}
\label{G*=}
\Gamma^*_{\underline{\alpha}\underline{\beta}}=-\left( \begin{matrix}0\;\; & \delta_{{\alpha}}{}^{\beta} \cr \delta{}^{{\alpha}}{}_{\beta} & 0  \end{matrix}\right)\; .    \qquad
\end{eqnarray}

Notice that there is another, more 'algorithmic' way to arrive at (\ref{SM0=SD0}). To this end we observe that, with the same assumption but allowing $\rho^\#$ to depend on $\tau$,  the variation of the action
(\ref{SM0=10+1=}) with respect to $X^*$ gives $d\rho^\#=0$. The solution of this equation is $\rho^\# =M=const$, (\ref{r++=M}) and  (\ref{SM0=SD0}) can be obtained by substituting this into (\ref{SM0=10+1=}).

Of course, the substitution of a dynamical equation back into the action is not an apparently consistent prescription, so that a better way to present the above described steps is to say that the dimensional reduction requires the momentum conjugate to the coordinate function corresponding to the reduced dimension  to be a constant.

Thus we have shown how the  dimensional reduction of the moving frame formulation of the M$0$ action produces D$0$-brane action.
Of course, the simplest reduction is achieved by setting the constant $L$-matrix equal to unity matrix
\begin{eqnarray}\label{L=1}
 L_{\underline{a}}{}^{\underline{b}}= (L_{\underline{a}}{}^b , L_{\underline{a}}{}^*)=  \delta_{\underline{a}}{}^{\underline{b}}\; .
\end{eqnarray}
Below, when considering dimensional reduction of spinor moving frame and of mM$0$ action, for simplicity  we will restrict ourselves by this case.

\setcounter{equation}0
\def\theequation{B.\arabic{equation}}

\section{Dimensional reduction of spinor moving frame: an embedding of $Spin(1,9)$ into $Spin(1,10)$}

In the previous Appendix A we have described the dimensional  reduction of the moving frame formulation of the M$0$-brane action to
D$0$-brane action without any use of spinor moving frame. However, our aim is to study the dimensional reduction of multiple M$0$-brane action and to this end the discussion of the dimensional reduction of spinor moving frame is inevitable.

D=11 spinor moving frame variables appropriate to the description of M$0$ and mM$0$ systems are rectangular blocks of the Spin(1,10) valued matrix
\begin{eqnarray}\label{harmV=11}
V_{\underline{\alpha}}^{(\underline{\beta})}= \left(\begin{matrix} v_{{\underline{\alpha}}{q}}^{\; +} , & v_{{\underline{\alpha}} q}^{\; -}
  \end{matrix}\right) \in Spin(1,10)\;  \qquad
\end{eqnarray}
which is called spinor moving frame matrix. This is related to the 11D moving frame (\ref{Uab=in11}) by the conditions of the Lorentz invariance of Dirac and charge conjugation matrices,
\begin{eqnarray}\label{VGVT=UG}
V^T\tilde{\Gamma}_{\underline{b}} V =  U_{\underline{b}}^{(\underline{a})} \tilde{\Gamma}_{(\underline{a})}\; , \qquad  \label{VTGV=UG} V
{\Gamma}^{(\underline{a})}  V^T = {\Gamma}^{\underline{b}} U_{\underline{b}}^{(\underline{a})}  \; , \qquad \\
\label{VCV=C}
V^TCV=C \; .   \qquad
\end{eqnarray}
Eq. (\ref{VCV=C}) allows to construct the elements of the inverse spinor moving frame matrix, which obey
\begin{eqnarray}\label{v-qv+p=}
& v^{-{\underline{\alpha}}}_q   v_{{\underline{\alpha}}}^{+{p}}=\delta_{qp}\;  , \qquad &
 v^{-{\underline{\alpha}}}_q   v_{{\underline{\alpha}}}^{-{q}}=0\;  , \qquad \nonumber \\
& v^{+{\underline{\alpha}}}_{{q}}  v_{{\underline{\alpha}}}^{+{p}}=0\;  , \qquad &
v^{+{\underline{\alpha}}}_{{q}} v_{{\underline{\alpha}}}^{-{p}} = \delta_{qp}
 \; ,  \qquad
\end{eqnarray}
in terms of the same spinor moving frame variables,
\begin{eqnarray}
\label{V-1=CV-A}   v_{q}^{\pm {\underline{\alpha}}}  =  \pm i C^{\underline{\alpha}\underline{\beta}}v_{\underline{\beta} q}^{\; \pm  }    \, . \qquad
 \end{eqnarray}

With a suitable Gamma matrix representation, Eqs. (\ref{VTGV=UG}) can be split into
\begin{eqnarray} \label{Iu--=vGv}
 v_{\underline{\alpha}q}^{\; -} \tilde{\Gamma}^{\underline{a}\underline{\alpha}\underline{\beta}} v_{\underline{\beta}p}^{\;-}= \delta_{qp}
U^{=}_{ \underline{a}} \; ,   \qquad \\
\label{M0:v+v+=u++}
 v_{q}^+ \tilde{\Gamma}_{ \underline{a}} v_{p}^+ = \; U_{ \underline{a}}^{\# } \delta_{qp}\; , \qquad
 v_{q}^-\tilde{\Gamma}_{ \underline{a}} v_{p}^+ = U_{ \underline{a}}^{i} \gamma^i_{qp}\; , \qquad
\end{eqnarray}
and
\begin{eqnarray}\label{M0:u--G=v-v-}
 2v_q^{-\underline{\alpha}} v_q^{-\underline{\beta}}= U^{=}_{\underline{a}}
\tilde{\Gamma}^{\underline{a}\underline{\alpha}\underline{\beta}} \; , \qquad
\\ \label{M0:u++G=v+v+}
 2 v_{q}^{+ \underline{\alpha}}v_{q}^{+}{}^{ \underline{\beta}}= \tilde{\Gamma}^{ \underline{a} \underline{\alpha} \underline{\beta}} U_{
 \underline{a}}^{\# }\; , \qquad
 2 v_{q}^{-( \underline{\alpha}}\gamma^i_{qp} v_p^{+\underline{\beta} )}=-  \tilde{\Gamma}^{ \underline{a} \underline{\alpha} \underline{\beta}}
 U_{ \underline{a}}^{i}\; . \qquad
\end{eqnarray}

Spinor moving frame can be used to split the single 11D Majorana spinor fermionic supervilebein form of 11D superspace (see (\ref{Eua=})) into two
16 component fermionic forms with SO(9) spinor indices and opposite SO(1,1) weights,
\begin{eqnarray}
\label{E-q=}
\underline{E}{}^{-q}= \underline{E}{}^{\underline{\alpha}}v_{\underline{\alpha} q}^{\; -}\; , \qquad \underline{E}{}^{+q}= \underline{E}{}^{\underline{\alpha}}v_{\underline{\alpha} q}^{\; +}\; . \qquad
 \end{eqnarray}

Our problem now is to find the expressions of the above 11D  spinor moving frame variables in terms of
$Spin(1,9)/Spin(9)$ spinor moving frame variables (\ref{harmV=10D0}),
\begin{eqnarray}\label{harmV=10}
v_\alpha{}^q \in Spin(1,9)\;  \qquad
\end{eqnarray} and its inverse obeying
\begin{eqnarray}\label{harmV-1=10}
v_q{}^\alpha v_\alpha{}^p=\delta_q{}^p \qquad \Leftrightarrow \qquad
 v_\alpha{}^q v_q{}^\beta= \delta_\alpha{}^\beta
\; . \qquad
\end{eqnarray}
This  corresponds to (provides  a square root of) the expression (\ref{U=Lu}) of 11D moving frame in terms of 10D moving frame with simplest choice
(\ref{L=1}).

The embedding of $Spin(1,9)$ group into $Spin(1,10)$ which defines such a dimensional reduction of the 11D spinor moving frame variables is defined by
\begin{eqnarray}\label{hV11=hv10}
 v_{{\underline{\alpha}} q}^{\; +}
=\frac 1 {\sqrt{2}} \left(\begin{matrix} v_{{\alpha}}^{\; q} \cr  -v_q{}^\alpha
  \end{matrix}\right)\; , \qquad  v_{{\underline{\alpha}} q}^{\; -}
=\frac 1 {\sqrt{2}} \left(\begin{matrix} v_{{\alpha}}^{\; q} \cr v_q{}^\alpha
  \end{matrix}\right)\; , \qquad
\end{eqnarray}
and complementary relations
 \begin{eqnarray}\label{hV-1=hv10}
 v^{+\underline{\alpha}}_q
=\frac 1 {\sqrt{2}} \left(\begin{matrix}  v_q{}^\alpha \cr v_{{\alpha}}^{\; q}
  \end{matrix}\right)\; , \qquad  v^{-\underline{\alpha}}_q
=\frac 1 {\sqrt{2}} \left(\begin{matrix}  v_q{}^\alpha \cr -v_{{\alpha}}^{\; q}
  \end{matrix}\right)\; . \qquad
\end{eqnarray}

\setcounter{equation}0
\def\theequation{C.\arabic{equation}}
\section{ Dimensional reduction of  mM$0$  and its comparison with mD$0$ action }

\subsection{Action for multiple M$0$-brane system}

The action for multiple M-wave (mM$0$) system proposed in
\cite{Bandos:2012jz} reads
\begin{eqnarray}
\label{SmM0=}  S_{mM0} &=& \int_{W^1} \rho^{\#}\, \underline{E}^{=} + \qquad \nonumber  \\ &&
+ \frac 1 {\mu^6} \int_{W^1} (\rho^{\#})^3\, \left(  tr\left(- {\bb P}^i D {\bb X}^i + 4i { \Psi}_q D
{\Psi}_q  \right) + \underline{E}^{\#} {\cal H} \right)+ \quad \nonumber \\ &&  +   \frac 1 {\mu^6}  \int_{W^1}
(\rho^{\#})^3\,  \underline{E}{}^{+q}  tr\left(4i (\gamma^i {\Psi})_q  {\bb P}^i + {1\over 2}
(\gamma^{ij} {\Psi})_q  [{\bb X}^i, {\bb X}^j]  \right) , \qquad
  \end{eqnarray}
where $\mu$ is a constant (of dimension of mass, $[\mu]=M$),  $ \underline{E}^{=} $, $\underline{E}^{\#}$ and $\underline{E}^{+q}$ are defined in (\ref{E--=Eu--}) and (\ref{E-q=}), $\rho^{\#}=\rho^{\#}(\tau)$ is the auxiliary worldline field which we have already met in the case of single M$0$-brane, ${\bb P}^i$, ${\bb X}^i$, ${ \Psi}_q$ are the bosonic and fermionic matrix fields describing SYM model (see sec. 3)  on the center of energy worldline of the mM$0$ system, and ${\cal H}$ is the SYM Hamiltonian  (\ref{HSYM=1}).
 The covariant derivatives of the matrix fields
 \begin{eqnarray}
\label{DXi=} D{\bb X}^i  &:=& d{\bb X}^i  + 2\Omega^{(0)} {\bb X}^i  - \underline{\Omega}^{ij} {\bb
X}^j+ [{\bb A},    {\bb X}^i] \; , \qquad \\ \label{DPsi:=} D\Psi_q  &:=& d\Psi_q
 + 3\Omega^{(0)} \Psi_q   -{1\over 4} \underline{\Omega}^{ij} \gamma^{ij}_{qp} {\Psi}_p+ [{\bb A},
 \Psi_q ] \;  \qquad
  \end{eqnarray}
 include the 1d gauge field  ${\bb A}=d\tau {\bb A}_\tau$ as well as Cartan forms constructed from the elements of 11D moving frame vectors,
 \begin{eqnarray}
\label{mM0:Om0=} \Omega^{(0)}= {1\over 4} U^{={\underline{a}}}dU_{\underline{a}}^{\#}\; , \qquad \underline{\Omega}^{ij}=
U^{i{\underline{a}}}dU_{\underline{a}}^{j}\;  . \qquad
  \end{eqnarray}
Finally
\begin{eqnarray}
\label{E+q=}
\underline{E}{}^{-q}= \underline{E}{}^{\underline{\alpha}}v_{\underline{\alpha} q}^{\; -}\; , \qquad \underline{E}{}^{+q}= \underline{E}{}^{\underline{\alpha}}v_{\underline{\alpha} q}^{\; +}\;  \qquad
 \end{eqnarray}
are projections of pull-back of the 11D fermionic supervielbein form onto the 11D spinor moving frame  (see (\ref{E-q=})).

The mM$0$ action (\ref{SmM0=}) is invariant under the local worldsheet supersymmetry
\begin{eqnarray}
\label{susy-thM0} \delta_\varepsilon {\Theta}^{\underline{\alpha}} &=& \varepsilon^{+q} (\tau)
v_q^{-\underline{\alpha}} \; , \quad
 \\
\label{susy-xM0} \delta_\varepsilon \hat{x}^{\underline{a}} &=& - i \hat{\theta} \Gamma^{\underline{a}}\delta_\varepsilon
\hat{\theta} +   {1\over 2}   U^{\underline{a}\#} i_\varepsilon \underline{E}^{=}\; , \qquad
\\
\label{susy-rhoM0}   \delta_\varepsilon \rho^{\#} &=& 0\; , \qquad \\
\label{susy-vM0}
 \delta_\varepsilon  v_q^{\pm{\underline{\alpha}}}&=&0
  \;   \Rightarrow  \quad  \delta_\varepsilon
  U_{\underline{a}}^{=}= \delta_\varepsilon U_{\underline{a}}^{\#}= \delta_\varepsilon U_{\underline{a}}^{i}=0\;
 ,  \qquad
\\
\label{susy-XM0}  \delta_\varepsilon {\bb X}^i   &=& 4i \varepsilon^{+} \gamma^i  \Psi \; , \quad
\delta_\varepsilon {\bb P}^i   = [(\varepsilon^{+} \gamma^{ij}  \Psi),  {\bb X}^j]\; ,\qquad \\
\label{susy-PsiM0}  \delta_\varepsilon \Psi_q &=&  {1\over 2} (\varepsilon^{+} \gamma^i)_q  {\bb
P}^i-  {i\over 16} (\varepsilon^{+} \gamma^{ij})_q  [{\bb X}^i, {\bb X}^j]\; ,\qquad \\
\label{susy-AM0}
 && \delta_\varepsilon {\bb A} = -  \underline{E}^{\#} \varepsilon^{+q}  \Psi_q + \underline{E}^{+}\gamma^i
 \varepsilon^{+}\;    {\bb X}^i
 \; ,  \qquad
\end{eqnarray}
where
\begin{eqnarray}
\label{iE==}
& i_\varepsilon \underline{E}^{=}= \frac {6}{\mu^6} (\rho^{\#})^2 tr \left(i {\bb P}^i\varepsilon^{+}
 \gamma^{i}\Psi
 -  {1\over 8} \varepsilon^{+} \gamma^{ij}\Psi  [{\bb X}^i, {\bb X}^j] \right) .   \qquad
 \end{eqnarray}

 \subsection{Dimensional reduction of mM$0$ action and its differences with mD$0$ action}

As in the case of single M$0$-brane, the dimensional reduction of the mM$0$ action implies the reduction of moving frame variables
by using (\ref{U--=Lu-L}) and  (\ref{U++=Lu+L}) with (\ref{L=1}). Then
\begin{eqnarray}
\label{dU++=} dU^\#_{\underline{a}}= dU^=_{\underline{a}}= \delta_{\underline{a}}^c du_c^{0}\qquad and
\qquad dU^i_{\underline{a}}= \delta_{\underline{a}}^c du_c^{i}
\qquad
  \end{eqnarray}
 so that the first of 11D Cartan forms (\ref{mM0:Om0=})
vanishes, $\Omega^{(0)}=0$, and $\underline{\Omega}^{ij}$ becomes identical to its 10D counterpart ${\Omega}^{ij}$.
As a result, the covariant derivatives of the matrix fields (\ref{DXi=}) and (\ref{DPsi:=}) coincide with the covariant derivatives  (\ref{DXi=10D}) and (\ref{DPsi:=10D}) used in the mD$0$ action (\ref{SmD0=}).

Now we can identify the matrix fields of mM$0$ and mD$0$ models as both of them are describing 1d reduction of
10D SYM model living on some worldline,
$$
\begin{matrix}{\bb X}^i & \mapsto & {\bb X}^i\; , \cr
 {\bb P}^i & \mapsto &  {\bb P}^i \;,\cr \Psi_q & \mapsto &  \Psi_q \; ,\cr
 {\bb A} & \mapsto & {\bb A}\; .
 \end{matrix}
$$
It is also natural to make the identification of  10 of 11D bosonic supervielbein forms
(\ref{Eb=ELb=}) with 10D supervielbein forms. With the simplest choice (\ref{L=1}) we have
\begin{eqnarray}
\label{uEb=Eb}
\underline{E}^{\underline{a}} \delta_{\underline{a}}{}^b:=  \underline{E}^b \mapsto  E^b\; .
\qquad
  \end{eqnarray}
This can be achieved by identification of the coordinate functions
$$\underline{x}^{\underline{a}} \delta_{\underline{a}}{}^b:=  \underline{x}^b \mapsto x^b \qquad and \qquad {\Theta}^{\underline{\alpha}} \mapsto  (\theta^{\alpha 1} \; \theta_\alpha^2)\; . $$
Besides these, in the 11D model we have eleventh bosonic coordinate function $X^*$ which enters the Cartan form
\begin{eqnarray}
\label{uE*=}
\underline{E}^* =dX^* - i \Theta \Gamma^* \Theta = dX^* +id\theta^{\alpha 1}\, \theta_{\alpha}^2- i\theta^{\alpha 1}\, d\theta_{\alpha}^2\; .
\qquad
  \end{eqnarray}

Taking into account the relation of moving frame variables,  (\ref{U--=Lu-L}) and  (\ref{U++=Lu+L})
with (\ref{L=1}), and
of the spinor moving frame variables, (\ref{hV11=hv10}), we find that the above identification and reduction rules imply that
\begin{eqnarray}
\label{11E--=}
\underline{E}^= &\mapsto & E^0- \underline{E}^* \; , \qquad \\
\label{11E++=}
\underline{E}^\# &\mapsto & E^0+ \underline{E}^* \; , \qquad \\
\label{11Eq+=}
\underline{E}^{q+} &\mapsto & \frac 1 {\sqrt{2}}(E^{q1}-E_q^2)   \; .
\qquad
  \end{eqnarray}


Let us follow the terms with pull-backs of the bosonic supervielbein forms in the mM$0$ action (\ref{SmM0=}):
\begin{eqnarray}
\label{SmM0=E0+E*+}
S_{mM0}= \int_{W^1}  E^0 \left( \rho^{\#}+ \frac {(\rho^{\#})^3} {\mu^6} {\cal H} \right)  + \int_{W^1}  E^* \left( -\rho^{\#}+ \frac {(\rho^{\#})^3} {\mu^6} {\cal H} \right) + \ldots \; ,
\qquad
  \end{eqnarray}
where $E^*$ has the form of Eq. (\ref{uE*=}) and ${\cal H}$ is defined in (\ref{HSYM=1}).

As we discussed in the case of single M0-brane, the mechanism of dimensional reduction can be formulated as setting to constant the momentum conjugate to additional coordinate field $X^*$ in $E^*$ of (\ref{E*=}). In our case, as $X^*$ enters the action only through its derivative and only linearly, this prescription can be formulated as obtaining the equations of motion for $X^*$,
$$
d\left( -\rho^{\#}+ \frac {(\rho^{\#})^3} {\mu^6} {\cal H} \right)=0\; , \qquad
$$
solving them by
\begin{eqnarray}
\label{rho=M+}
\rho^{\#}-\frac {(\rho^{\#})^3} {\mu^6} {\cal H} =M
\qquad
  \end{eqnarray}
with some constant $M$, and substituting the result back into the action.

The problem with such an action is that it includes $\rho^{\#}$ which is a nonlinear function of the relative motion variables defined by a solution of Eq. (\ref{rho=M+}) with ${\cal H}$ from (\ref{HSYM=1}).
For large ${\cal H}$ one can use an explicit expression for  $\rho^{\#}=\rho^{\#}({\cal H})$ obtained from the  Cardano formula,
$$
\rho^{\#}= \frac {\mu^2 M^{1/3}}{(2{\cal H})^{1/3}} \left(\left(\sqrt{1-\frac {4\mu^6}{27{\cal H}M^2}} -1 \right)^{1/3}- \left(\sqrt{1-\frac {4\mu^6}{27{\cal H}M^2}} +1 \right)^{1/3}\right)\; , \qquad
$$
but this is not too suggestive. It is more practical to keep $\rho^{\#}=\rho^{\#}({\cal H})$ implicit, as a solution of (\ref{rho=M+}), but use the variation of $\rho^{\#}$ which preserves (\ref{rho=M+})
and hence is expressed in terms of variation of the SYM hamiltonian ${\cal H}$ (\ref{HSYM=1}) by
\begin{eqnarray}
\label{vrho=}
\delta \rho^{\#} = \left(1-\frac {3(\rho^{\#})^2} {\mu^6} {\cal H}\right)^{-1} \; \frac {(\rho^{\#})^3} {\mu^6} \delta {\cal H}\; .
\qquad
  \end{eqnarray}

Using (\ref{rho=M+}) and keeping implicit its solution $\rho^{\#}=\rho^{\#}({\cal H})$, we can write Eq. (\ref{SmM0=E0+E*+}) in the form
\begin{eqnarray}
\label{SmM0=2r-ME0+WZ}
S_{mM0}= \int_{W^1}  E^0 \left(2 \rho^{\#}({\cal H})-M\right)  -iM \int_{W^1}  \left(d\theta^{\alpha 1}\, \theta_{\alpha}^2- \theta^{\alpha 1}\, d\theta_{\alpha}^2\right) + \ldots \; .
\qquad
  \end{eqnarray}

The explicitly written terms are clearly different from the first line of (\ref{SmD0=}), were in the first term
$ E^0 $ is multiplied just by the constant $M$.
Neither they are related with Panda-Sorokin action  (\ref{S=PS}). To make this explicit, let us write the moving frame formulation of the  Panda-Sorokin action:
\begin{eqnarray}
\label{S'=PS}
 S'_{PS} &=&\int_{{\cal W}^1} E^0 {\cal M}({\bb X}, {\bb P},\Psi)  - i
  \int_{{\cal W}^1} {\cal M}({\bb X}, {\bb P},\Psi) (d{\theta}^{\alpha 1}\,{\theta}_{\alpha}^{2} - {\theta}^{\alpha 1}d{\theta}_{\alpha}^{2} ) + \quad\nonumber \\ &&
 +  \int_{{\cal W}^1} Tr\left(- {\bb P}^i d{\bb X}^i + 4i\Psi_qd\Psi_q \right) \; .  \qquad
\end{eqnarray}
Both terms in the first line of this equation involve the same function of the internal matrix variables,
${\cal M}({\bb X}, {\bb P},\Psi)$, which is not the case for the terms in (\ref{SmM0=2r-ME0+WZ}).

To conclude, the discussion of the dimensional reduction of the 11D mM$0$ action (\ref{SmM0=}) has resulted in the conclusion that,
besides that it is not easy to work with such an action which is non-linear in matrix fields, it does not look related neither to our mD$0$-brane system described by the much simpler  functional (\ref{SmD0=}), nor to Panda-Sorokin action (\ref{S=PS}) in its moving frame formulation  (\ref{S'=PS}).

\end{document}